\documentclass[pra,aps,showpacs,twocolumn,superscriptaddress]{revtex4}
\usepackage{amsfonts}
\usepackage{amsmath}
\usepackage{amssymb}
\usepackage{graphicx}
\usepackage{txfonts}

\begin{document}

\title{Quantum interferometry with binary-outcome measurements in the presence of phase diffusion}
\author{X. M. Feng}
\affiliation{Department of Physics, Beijing Jiaotong University, Beijing 100044, China}
\author{G. R. Jin}
\email{grjin@bjtu.edu.cn}
\affiliation{Department of Physics, Beijing Jiaotong University, Beijing 100044, China}
\author{W. Yang}
\email{wenyang@csrc.ac.cn}
\affiliation{Beijing Computational Science Research Center, Beijing 100084, China}
\date{\today }

\begin{abstract}
Optimal measurement scheme with an efficient data processing is important in quantum-enhanced interferometry. Here we prove that for a general binary-outcome measurement, the simplest data processing based on inverting the average signal can saturate the Cram\'{e}r-Rao bound. This idea is illustrated by  binary-outcome homodyne detection, even-odd photon counting (i.e., parity detection), and zero-nonzero photon counting that have achieved super-resolved interferometric fringe and shot-noise limited sensitivity in coherent-light Mach-Zehnder interferometer. The roles of phase diffusion are investigated in these binary-outcome measurements. We find that the diffusion degrades the fringe resolution and the achievable phase sensitivity. Our analytical results confirm that the zero-nonzero counting can produce a slightly better sensitivity than that of the parity detection, as demonstrated in a recent experiment.
\end{abstract}

\pacs{42.50.Ar, 42.25.Hz, 03.65.Ta}
\maketitle

\section{Introduction}

\begin{figure}[ptbh]
\begin{center}
\includegraphics[width=0.95\columnwidth]{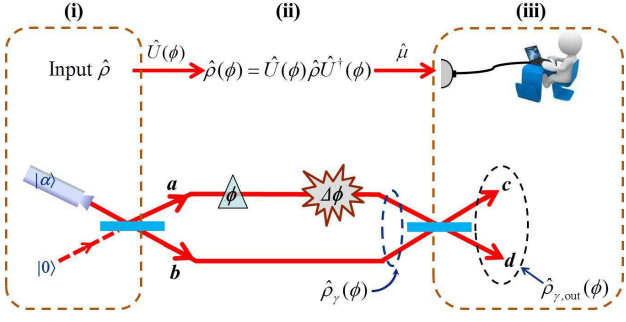}
\caption{(Color online) Three steps of quantum interferometry: (i) A probe state $\hat{\rho}$ is prepared and is injected into the interferometer; (ii) an unknown phase shift is accumulated during a unitary process $\hat{U}(\phi)$; (iii) the phase information is extracted via a detection $\hat{\mu}$ and a proper choice of data processing. A Mach-Zehnder interferometer fed with a coherent-state light is considered to investigate the role of phase diffusion (indicated by $\Delta\phi$) in the binary-outcome measurements.}
\label{fig1}
\end{center}
\end{figure}

Optical phase measurement in a Mach-Zehnder interferometer (MZI) consists of three steps [see e.g., Refs.~\cite{Kay,Helstrom}, and also Fig.~\ref{fig1}]. First, a probe state $\hat{\rho}$ is prepared and is injected into the MZI. Second, it undergoes a dynamical process described by a unitary operator $\hat{U}(\phi )$ and evolves into a phase-dependent state $\hat{\rho}(\phi )=\hat{U}(\phi )\hat{\rho}\hat{U}^{\dagger }(\phi )$, where $\phi $ is a dimensionless phase shift. Finally, a detection $\hat{\mu}$ and a specific data processing are made at the output ports to obtain an interferometric signal $\langle \hat{\mu}(\phi )\rangle =\mathrm{Tr}[\hat{\rho}(\phi )\hat{\mu}]$, which  shows  an  oscillatory
pattern with  the  resolution  determined  by  the  full width  at  half maximum  $(\mathrm{FWHM})$  of  the
signal and the wavelength $\lambda$, i.e., $\Delta x\propto \mathrm{FWHM}\times \lambda /(2\pi )$~\cite{Born,Boto}. In addition, the phase sensitivity is determined by the error-propagation formula $\delta \phi =\Delta\hat{\mu}/|\partial \langle \hat{\mu}\rangle /\partial \phi |$, with the square root of the variance $\Delta \hat{\mu}\equiv \sqrt{\langle \hat{\mu}^{2}\rangle-\langle \hat{\mu}\rangle ^{2}}$~\cite{Bevington}.

An optimal measurement scheme with a proper choice of data processing is important to improve the resolution and the sensitivity~\cite {Bevington,Dowling}. For instance, in a coherent-light MZI, the intensity measurement at one of the two output ports produces the signal $\langle \hat{\mu}\rangle \propto \sin ^{2}(\phi /2)$ or $\cos ^{2}(\phi /2)$, which exhibits the $\mathrm{FWHM}=\pi $ and hence the fringe resolution $\Delta x\sim \lambda /2$, known as the Rayleigh resolution limit~\cite{Born,Boto}. Resch \textit{et al.}~\cite{Resch} demonstrated that coherent light can provide a better resolution beyond the Rayleigh limit (i.e., super-resolution); however, the achievable sensitivity is much worse than the shot-noise limit $1/\sqrt{N}$, where $N$ is average number of photons. Pezz\'{e} \textit{et al.}~\cite{Pezze} have proposed that coincidence photon counting with a Bayesian estimation strategy results in the shot-noise limited sensitivity over a broad phase interval. However, the visibility of the coincidence rates decays quickly with the increased number of photons being detected~\cite{Afek2010}.

The parity measurement gives binary outcomes, dependent upon even or odd number of photons at one of two output ports. It originates from atomic spectroscopy with an ensemble of trapped ions~\cite{Bollinger} and was discussed in the context of optical interferometry by Gerry~\cite{Gerry2} and subsequently by others~\cite{Anisimov,Seshadreesan}. Recently, Gao \textit{et al.}~\cite{Gao} proposed that the parity measurement can lead to the super-resolution in the coherent-light MZI. Using a binary-outcome homodyne detection, Distante \textit{et al.}~\cite{Andersen} have demonstrated a super-resolution with the $\mathrm{FWHM}\sim\pi/\sqrt{N}$ and a phase sensitivity close to the shot-noise limit. Most recently, Cohen \textit{et
al.}~\cite{Eisenberg} have realized the parity measurement in a polarization version of the MZI. In contrast to the previous theory~\cite{Gao}, they found that the peak height of signal decreases as the average photon number $N$ increases, which in turn leads to divergent phase sensitivity at certain phase shifts. More surprisingly, they found that the zero-nonzero photon counting (hereinafter, called the $Z$ detection) can saturate the shot-noise limit and gives a slightly better sensitivity than that of the parity measurement. Since both the parity and the $Z$ detections are simply two kinds of photon counting, the reason the $Z$ detection prevails is still lacking.

In this paper, we present a unified description to the above coherent-light MZI experiments~\cite{Andersen,Eisenberg}, using general expressions of conditional probabilities for detecting an outcome in homodyne detection and in photon counting measurement. We first show that for a general binary-outcome measurement, the simplest data processing based on inverting the average signal can saturate the Cram\'{e}r-Rao (CR) bound. For such measurements, more complicated data processing techniques such as maximal likelihood estimation or Bayesian estimation are not necessary. This conclusion is independent of the input states and the presence of noises. Next, we investigate the role of phase diffusion~\cite{Qasimi,Teklu,Liu,Brivio,Genoni11,Genoni12,Escher12,Zhong,Bardhan} on the binary-outcome homodyne detection, the parity measurement, and the $Z$ measurement. Our analytical results show that the diffusion plays a role in a form of $N\gamma$, rather than the phase-diffusion rate $\gamma$ and the mean photon number $N$ alone. When $N\gamma \ll 1$, the effect of phase noise can be negligible; both the resolution and the best sensitivity almost follow the shot-noise scaling $\sim 1/\sqrt{N}$. As $N\gamma$ increases, both the resolution and the sensitivity deviate from the scaling. Analytically, we confirm that the $Z$ detection gives a better sensitivity than that of the parity detection, as demonstrated recently by Cohen~\textit{et al.}~\cite{Eisenberg}.

\section{Quantum phase measurements with binary outcomes}

We first briefly review quantum phase measurement in a standard MZI fed with coherent-state light~(see Fig.~\ref{fig1}). Similar to the experimental setup~\cite{Andersen}, a coherent state $|\alpha \rangle $ with amplitude $\alpha =\sqrt{N}$ is injected into one port of the MZI and the other port is left in vacuum $|0\rangle $. After a 50:50 beamsplitter~\cite{Bs,Gerrybook}, the photon state becomes a product of coherent states $|\alpha /\sqrt{2}\rangle _{a}\otimes |\alpha /\sqrt{2}\rangle _{b}$, where the subscript $a$ ($b$) denotes the path or the polarization mode. Second, the phase shift $\phi $ is accumulated in one arm of the interferometer~\cite{Andersen} through an unitary evolution $\hat{U}(\phi )=\exp (-i\phi\hat{N}_{a})$, with the number operator $\hat{N}_{a}=\hat{a}^{\dag }\hat{a}$. The phase accumulation results in the photon state $\hat{\rho}(\phi )=\hat{\rho}_{a}(\phi )\otimes \hat{\rho}_{b}(0)$, where the density operators for the two modes are given by $\hat{\rho}_{a}(\phi)\equiv \hat{U}(\phi )\hat{\rho}_{a}(0)\hat{U}^{\dag }(\phi )=|\alpha e^{-i\phi }/\sqrt{2}\rangle_{aa}\langle \alpha e^{-i\phi}/\sqrt{2}|$ and $\hat{\rho}_{b}(0)=|\alpha /\sqrt{2}\rangle _{bb}\langle \alpha /\sqrt{2}|$, respectively. After the second 50:50 beamsplitter, the phase-encoded state becomes
\begin{equation}
\hat{\rho}_{\mathrm{out}}(\phi )=\hat{B}_{1/2}\hat{\rho}(\phi )\hat{B}_{1/2}^{\dag }=|\psi _{\mathrm{out}}(\phi )\rangle \langle \psi _{\mathrm{out}}(\phi )|,  \label{output0}
\end{equation}
where $\hat{B}_{1/2}$ denotes the beam-splitter operator~\cite{Bs,Gerrybook}, and $|\psi _{\mathrm{out}}\rangle =|\alpha (e^{-i\phi }-1)/2\rangle_{c}\otimes |\alpha (e^{-i\phi }+1)/2\rangle _{d}$ is the output state for the optical modes $c$ and $d$. Finally, a measurement and data processing are performed to obtain phase-sensitive output signal, which determine the fringe resolution and the phase sensitivity.

For instance, a homodyne detection of the phase quadrature $\hat{p}=(\hat{c}-\hat{c}^{\dag })/2i$ at the output port $c$ is described by the projection operators $\{|p\rangle \langle p|\}$ with $\hat{p}|p\rangle =p|p\rangle $. The probability for detecting an outcome $p$ is simply given by $P(p|\phi)\equiv \mathrm{Tr}[\hat{\rho}_{\mathrm{out}}(\phi )|p\rangle \langle p|]$, with its explicit form,
\begin{equation}
P(p|\phi )=\sqrt{\frac{2}{\pi }}\,{\exp }\left[ {-2}\left( {p+\frac{\sqrt{N}{\sin \phi }}{2}}\right) ^{2}\right] ,  \label{homodyne}
\end{equation}
where we have used the wave function of a coherent state $\langle p|\alpha\rangle \equiv (2/\pi )^{1/4}\exp [-(p-y_{0})^{2}-2ix_{0}p+ix_{0}y_{0}]$, with $x_{0}=\mathrm{Re}(\alpha )$ and $y_{0}=\mathrm{Im}(\alpha )$. The output signal is then given by $\langle \hat{p}(\phi )\rangle=\int_{\mathbb{R}}\!dppP(p|\phi )\propto \sqrt{N}\sin \phi $, which exhibits the $\mathrm{FWHM}\approx \pi $ and hence the Rayleigh limit in fringe resolution. The Fisher information of the homodyne measurement is given by
\begin{equation}
F(\phi )=\int_{\mathbb{R}}\!\!dp\frac{1}{P(p|\phi )}\left[ \frac{\partial P(p|\phi )}{\partial \phi }\right] ^{2}=N\cos ^{2}\phi ,  \label{FH}
\end{equation}
which yields the CR bound $\delta \phi _{\mathrm{CRB}}=1/(\sqrt{N}|\cos\phi |)$, dependent upon the true value of phase shift. Only at $\phi _{\min}=k\pi $ for integers $k$, the lower bound of phase sensitivity can reach the shot-noise limit.

Next, let us consider a general photon counting characterized by a set of projection operators $\{|n,m\rangle \langle n,m|\}$, with the two-mode Fock states $|n,m\rangle \equiv |n\rangle _{c}\otimes |m\rangle _{d}$. The probability for detecting $n$ photons at the output port $c$ and $m$ photons at the port $d$, i.e., the coincidence rate $P(n,m|\phi )\equiv \langle n,m| \hat{\rho}_{\mathrm{out}}(\phi )|n,m\rangle $ \cite{Afek2010}, is given by
\begin{equation}
P(n,m|\phi )=\frac{e^{-N}}{n!m!}\left( N\sin ^{2}\frac{\phi }{2}\right)^{n}\left( N\cos ^{2}\frac{\phi }{2}\right) ^{m},  \label{coincidence}
\end{equation}
with the mean photon number $N=\alpha ^{2}$. For a light intensity measurement at the output port $d$, we obtain the signal that is proportional to $\langle \hat{N}_{d}(\phi )\rangle =\sum_{m}mP(m|\phi)=N\cos ^{2}(\phi /2)$, where $\hat{N}_{d}=\hat{d}^{\dag }\hat{d}$ and the probability $P(m|\phi )=\sum_{n}P(n,m|\phi )$. It is easy to find that the $\mathrm{FWHM}=\pi $ and the resolution $\Delta x\propto \lambda /2$, as discussed before. In addition, we obtain the Fisher information of this light-intensity detection,
\begin{equation}
F(\phi )=\sum_{m}\frac{1}{P(m|\phi )}\left[ \frac{\partial P(m|\phi )}{\partial \phi }\right] ^{2}=N\sin ^{2}\frac{\phi }{2},  \label{Fisher}
\end{equation}
and hence the lower bound $\delta \phi _{\mathrm{CRB}}=1/[\sqrt{N}|\sin(\phi /2)|]$, which reaches the shot-noise limit at $\phi _{\min }=(2k+1)\pi$ for integers $k$.

From Eq.~(\ref{coincidence}), one can note that the coincidence rate with $ nm\neq 0$ shows multifold oscillations as a function of $\phi $, leading to an enhanced phase resolution beyond the Rayleigh limit. As demonstrated by Afek \textit{et al}.~\cite{Afek2010}, however, the visibility of the multifold oscillations decays quickly with the increased number of photons being detected. Using path-entangled NOON states, the super-resolution of interferometric fringe $\Delta x\propto \lambda /(2N)$ is possible~\cite{Mitchell,Walther,Chen}, but with $N\lesssim 5$~\cite{Afek}.

The homodyne detection and the photon counting with a proper choice of data processing can improve the phase resolution. Recently, it has been shown that the binary-outcome homodyne detection~\cite{Andersen} and photon counting~\cite{Eisenberg} in the coherent light MZI lead to the super-resolution $\Delta x\propto \lambda /(2\sqrt{N})$ and the sensitivity close to the shot-noise limit. We show below that the above observations can be understood by proper data processing to Eqs. (\ref{homodyne}) and (\ref{coincidence}). Moreover, we verify that the CR bound of \emph{any} binary-outcome measurements can be saturated by the simplest data processing based on inverting the average signal.

In quantum interferometry, any measurement of the Hermitian operator $\hat{\mu}$ with respect to arbitrary phase-encoded state $\hat{\rho}(\phi )$ can be modeled by projection onto the orthonormalized eigenstates $\{|\mu\rangle \langle \mu |\}$ of operator $\hat{\mu}$, with $\hat{\mu}|\mu\rangle =\mu |\mu \rangle $. Here, we focus on projection measurements with binary outcomes $\mu _{\pm }$. The associated probabilities are defined as $P(\pm |\phi )\equiv \mathrm{Tr}[\hat{\rho}(\phi )|\mu _{\pm }\rangle \langle \mu _{\pm }|]=\langle \mu _{\pm }|\hat{\rho}(\phi )|\mu _{\pm }\rangle $.
The simplest data processing is based on the average signal and its second moment, i.e., $\langle \hat{\mu}^{k}(\phi )\rangle \equiv \mu_{+}^{k}P(+|\phi )+\mu _{-}^{k}P(-|\phi )$ for $k=1$, $2$. Using the normalization condition $P(+|\phi )+P(-|\phi )=1$, we obtain the variance $ (\Delta \hat{\mu})^{2}=(\mu _{+}-\mu _{-})^{2}P(+|\phi )P(-|\phi )$. For phase-independent outcomes $\mu _{\pm }$, we further obtain the slope of signal $|\partial \langle \hat{\mu}\rangle /\partial \phi |=|(\mu _{+}-\mu_{-})\partial P(+|\phi )/\partial \phi |$. Therefore, the error-propagation formula gives the phase sensitivity
\begin{equation}
\delta \phi =\frac{\Delta \hat{\mu}}{|\partial \langle \hat{\mu}\rangle/\partial \phi |}=\frac{\sqrt{P(+|\phi )P(-|\phi )}}{|\partial P(+|\phi )/\partial \phi |}=\frac{1}{\sqrt{F\left( \phi \right) }}, \label{sensitivity}
\end{equation}
where the Fisher information for the binary-outcome detection is given by Eq.~(\ref{Fisher}) with $m=\pm $. Obviously, the sensitivity obtained from the error-propagation formula can saturate the CR bound over the entire phase interval. Actually, this conclusion holds not only for the projection measurements, but also for the most general kind of quantum measurement (i.e., positive operator-valued measure). In addition, Eq.~(\ref{sensitivity}) remains valid for arbitrary input state and is independent from the presence of noises. Previously, Seshadreesan \textit{et al.}~\cite{Seshadreesan} found that for the parity measurement, the inversion estimator can reach the CR bound. This measurement is a special case of photon counting at one of two output ports with the outcomes $\mu _{\pm }=\pm 1$ (see below).

\subsection{Binary-outcome homodyne detection}

\begin{figure}[ht]
\begin{center}
\includegraphics[width=0.9\columnwidth]{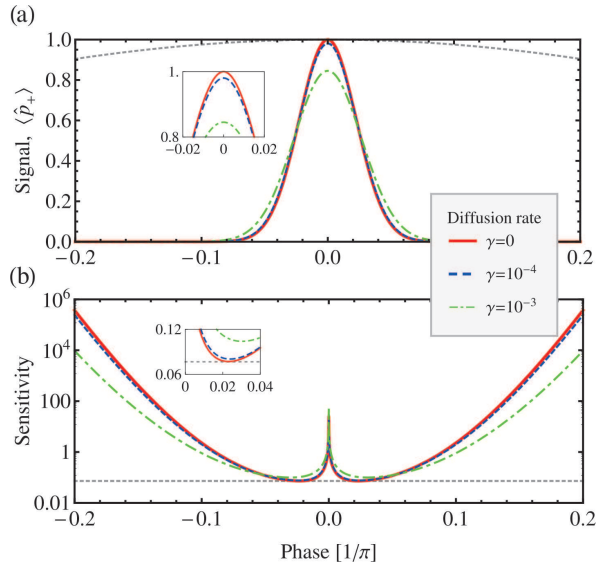}
\caption{(Color online) Normalized output signal (a) and phase sensitivity (b) for the binary-outcome homodyne detection with a fixed number of photons $N=200$ and various phase-diffusion rates $\gamma=0$ (red solid), $10^{-4}$ (blue dashed), and $10^{-3}$ (green dot-dashed). The signals in (a), normalized by $\sqrt{2/\pi}$, and the sensitivities in (b) are the exact results, obtained by numerically integrating Eq.~(\ref{rhoout-2HD}). Gray dotted line in (a) is the normalized signal of the intensity measurement, i.e., $\cos^{2}(\phi/2)$; while in (b), it indicates the best sensitivity $1.03/\sqrt{N}$, given by Eq.~(\ref{bestsensitivity-2HD}) for $\gamma=0$. Insets: zoomed output signal and the phase sensitivity near the phase origin $\phi=0 $. }
\label{fig2}
\end{center}
\end{figure}

Recently, Distante \textit{et al.}~\cite{Andersen} demonstrated a binary-outcome homodyne detection by dividing the total data of phase quadrature into binary outcomes: $|p|\leq p_{0}$ and $|p|>p_{0}$, denoted by \textquotedblleft $+$" and \textquotedblleft $-$", respectively, with the associated probabilities
\begin{equation*}
P(+|\phi )=\int_{-p_{0}}^{p_{0}}\!\!dpP(p|\phi )\text{, \ \ \ }P(-|\phi)=1-P(+|\phi ),
\end{equation*}
where $P(p|\phi )$ is given by Eq.~(\ref{homodyne}). Note that the conditional probability for detecting the outcome $+$ is the same to the interferometric signal ${\langle \hat{p}_{+}(\phi )\rangle =}\mathrm{Tr}[\hat{\rho}_{\mathrm{out}}{(\phi )}\hat{p}_{+}]$, with the observable $\hat{p}
_{+}=\int_{-p_{0}}^{p_{0}}\!dp|p\rangle \langle p|$ and $\hat{p}_{+}^{2}=\hat{p}_{+}$. Moreover, this kind of data processing results in a super-resolved fringe pattern~\cite{Andersen}, which can be understood by considering the limit $p_{0}\rightarrow 0$, corresponding to a detection of the phase quadrature with $p=0$. In this case, the observable becomes ${\hat{p}_{+}=}|p=0\rangle \langle p=0|$ and the interferometric signal is therefore given by ${\langle \hat{p}_{+}\rangle =}P(p=0|\phi )$, with its explicit form [cf. Eq.~(\ref{homodyne})]
\begin{equation}
{\langle \hat{p}_{+}(\phi )\rangle }=\,\sqrt{\frac{2}{\pi }}\exp \left( -\frac{N}{2}\sin ^{2}\left( \phi \right) \right) ,  \label{p-homody}
\end{equation}
which, as illuminated by the red solid line of Fig.~\ref{fig2}(a), becomes narrowing in a comparison with that of the intensity detection [$\propto \cos ^{2}(\phi /2)$, the gray dotted line]. To understand this behavior, we now analyze Eq.~(\ref{p-homody}) near the phase origin $\phi \sim 0$, and obtain ${\langle \hat{p}_{+}\rangle }\propto \exp (-N\phi ^{2}/2)$, which gives the $\mathrm{FWHM}=2\sqrt{(2\ln 2)/N}<\pi /\sqrt{N}$ and hence the resolution $\Delta x<\lambda /(2\sqrt{N})$. Clearly, the resolution is improved by a factor $\sqrt{N}$ beyond the Rayleigh limit $\lambda /2$~\cite{Andersen}.

According to the error-propagation formula, i.e., Eq.~(\ref{sensitivity}), we further obtain the phase sensitivity
\begin{equation}
\delta \phi _{H}=\frac{1}{\sqrt{F(\phi )}}=\frac{2}{N}\frac{\left( \sqrt{\frac{\pi }{2}}e^{(N/2)\sin ^{2}\phi }{-1}\right) ^{1/2}}{|\sin (2\phi)|},  \label{delphi}
\end{equation}
which saturates the CR bound. As shown by the red solid line of Fig.~\ref{fig2}(b), one can find that the sensitivity diverges at the phase origin $\phi =0$, due to the nonzero variance of ${\langle \hat{p}_{+}\rangle }$ and the vanishing slope of signal as $\phi \rightarrow 0$. One can also see this from Eq.~(\ref{delphi}). It is interesting to note that local minimum of $\delta \phi _{H}$ (i.e., the best sensitivity $\delta \phi _{H,\min }$) can be obtained by maximizing the slope $|\partial {\langle \hat{p}_{+}\rangle }/\partial \phi |=N|\sin (2\phi )|{\langle \hat{p}_{+}\rangle }/2$,  where $\langle \hat{p}_{+}\rangle $ is given by Eq.~(\ref{p-homody}). As a result, one can obtain the optimal phase shift $\phi _{\mathrm{\min }}$,  obeying $N\sin^{2}(2\phi _{\mathrm{\min }})=4\cos (2\phi _{\mathrm{\min }})$. In the large-$N$ limit, it gives $\exp [N\sin ^{2}(\phi _{\mathrm{\min }})/2]\approx\sqrt{e}$, due to $\phi _{\mathrm{\min }}\approx 0$ and $N\phi_{\mathrm{\min }}^{2}\approx 1$. Finally, from Eq.~(\ref{delphi}), we obtain the best sensitivity,
\begin{equation}
\delta \phi _{H,\min }=\frac{\left( \sqrt{\frac{\pi }{2}}e^{(N/2)\sin ^{2}\phi _{\mathrm{\min }}}-{1}\right) ^{1/2}}{|N\cos (2\phi _{\mathrm{\min }})|^{1/2}}\approx \frac{(\sqrt{e\pi /2}-{1)}^{1/2}}{\sqrt{N}},
\label{delmin_2HD}
\end{equation}
in agreement with Distante \textit{et al.}~\cite{Andersen}. For nonzero $p_{0}$ ($=1/2$), it has been demonstrated that the best sensitivity can reach $\delta \phi _{H,\min }\approx 1.37/\sqrt{N}$, approaching the shot-noise limit~\cite{Andersen}.

\subsection{Parity detection}

\begin{figure}[ptbh]
\begin{center}
\includegraphics[width=1\columnwidth]{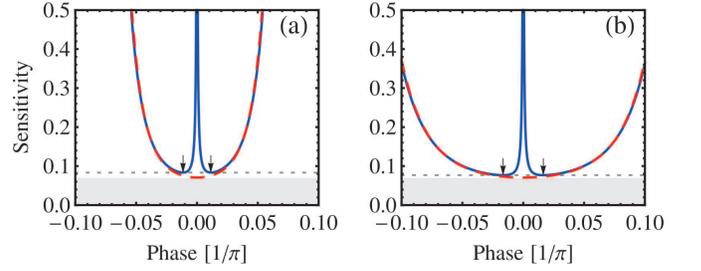}
\caption{(Color online) Phase sensitivities $\delta\phi$ against dimensionless phase shift $\phi$ (in units of $1/\pi$ ) for the parity detection (a) and the $Z$ detection (b). Solid (red dashed) lines: the exact solutions of $\delta\phi$ with (without) the phase diffusion for the mean photon number $N=200$ and the phase-diffusion rate $\gamma=10^{-4}$. Horizontal dotted lines: the best sensitivities, predicted by Eqs.~(\ref{sen_P}) and (\ref{sen_Z}). The arrows indicate the positions of the optimal phase shift $\phi_{\min}$. Shaded areas: the shot-noise limit $1/\sqrt{N}$ and below.}
\label{fig3}
\end{center}
\end{figure}

According to Gao \textit{et al.}~\cite{Gao}, the parity detection in the standard MZI also results in the super-resolution. The parity measurement~\cite{Bollinger,Gerry2,Anisimov,Seshadreesan} at the output port $c$, described by the parity operator $\hat{\Pi}=(-1)^{\hat{c}^{\dag}\hat{c}}$, groups the photon counting $\{n,m\}$ into binary outcomes $\pm 1$, according to even or odd number of photons $n$ at that port $c$. Such a kind of data processing provides an optimal phase estimator for the input path-symmetric states~\cite{Seshadreesan}. The conditional probabilities $P(\pm 1|\phi )$ are obtained by a sum of $P(n,m|\phi )$ over the even or the odd $n$'s, namely
\begin{equation*}
P(+1|\phi )=\sum_{m,n}^{\mathrm{even}\text{ }n}P(n,m|\phi )=\frac{1}{2}\left( 1+e^{-2N\sin ^{2}(\phi/2)}\right) ,
\end{equation*}
and $P(-1|\phi )=1-P(+1|\phi )$. In deriving the above result, we have used the identity $\sum_{n}^{\mathrm{even}\text{ }n}x^{n}/n!=\cosh x$. The signal corresponds to the expectation value of the parity operator $\hat{\Pi}$ with respect to $\hat{\rho}_{\mathrm{out}}$, given by
\begin{equation}
\langle \hat{\Pi}(\phi )\rangle =P(+1|\phi )-P(-1|\phi )=e^{-2N\sin^{2}(\phi /2)},  \label{P+}
\end{equation}
which coincides with Gao \textit{et al.}~\cite{Gao}. Near $\phi=0$, the signal $\langle \hat{\Pi}\rangle \approx \exp(-N\phi^{2}/2)$, similar to the binary-outcome homodyne detection of Eq.~(\ref{p-homody}), results in the super-resolution with $\Delta x\sim\lambda/(2\sqrt{N})$. Due to $\hat{\Pi}^{2}=1$, the phase sensitivity is given by
\begin{equation}
\delta \phi _{\Pi }=\frac{\sqrt{e^{4N\sin ^{2}(\phi/2)}-1}}{N\left\vert \sin \phi \right\vert }\approx \frac{1}{\sqrt{N}}\left( 1+\frac{1+2N}{8}\phi ^{2}\right) ,  \label{delphi_P}
\end{equation}
which saturates the CR bound over the entire phase interval. In the last step, the sensitivity is expanded up to the second order of $\phi$~\cite{Gao}. Obviously, the sensitivity can reach the shot-noise limit at $\phi =0$ [see the red dashed line of Fig.~\ref{fig3}(a)].

\subsection{$Z$ detection}

As the final example, we consider the zero-nonzero photon counting (named as the $Z$ detection) at the output port $c$, following Ref.~\cite{Eisenberg}. In this scheme, the counting data $\{n,m\}$ are classified into binary outcomes: $0$ for $n=0$ and $\emptyset $ for $n\neq 0$, with the associated probabilities
\begin{equation}
P(0|\phi )=\sum_{m}P(n=0,m|\phi )=e^{-N\sin ^{2}(\phi/2)}, \label{Zout}
\end{equation}%
and $P(\emptyset |\phi )=1-P(0|\phi )$. The output signal $\langle\hat{Z}(\phi)\rangle=P(0|\phi)$ corresponds to the expectation value of the observable $\hat{Z}=|0\rangle _{cc}\langle 0|$ with respect to $\hat{\rho}_{\mathrm{out}}$. Note that at $\phi =0$, the output state is indeed $|\psi_{\mathrm{out}}\rangle=|0\rangle_{c}\otimes |\alpha\rangle_{d}$, so no photon is detected at output port $c$ and $P(0|\phi =0)=\langle\hat{Z}(0)\rangle=1$. Near the phase origin, one can find that the signal $\langle \hat{Z}\rangle \approx \exp(-N\phi^{2}/4)$ and hence the $\mathrm{FWHM}=4\sqrt{\ln 2/N}\sim\pi/\sqrt{N}$, leading to a super-resolved fringe pattern. However, the resolution becomes worse by a factor $\sqrt{2}$ than that of the previous detections \cite{Eisenberg}. Using $\hat{Z}^{2}=\hat{Z}$ and the error-propagation formula, we obtain the phase sensitivity
\begin{equation}
\delta \phi _{Z}=\frac{2\sqrt{e^{N\sin ^{2}(\phi/2)}-1}}{N\left\vert \sin \phi \right\vert }\approx \frac{1}{\sqrt{N}}\left( 1+\frac{1+N/2}{8} \phi ^{2}\right) ,  \label{deltaphi_Z}
\end{equation}
which can also saturate the CR bound and reach the shot-noise limit at $\phi =0$, as shown by the red dashed line of Fig.~\ref{fig3}(b).

With the above binary-outcome detections, one can note that both the phase resolution and the best sensitivity exhibit the shot-noise scaling $\sim 1/
\sqrt{N}$. Specially, the parity and the $Z$ detections show the best sensitivity at $\phi =0$ due to the peak heights $\langle \hat{\Pi}(0)\rangle =\langle \hat{Z}(0)\rangle =1$. The finite value of $\delta \phi $ at $\phi =0$ (i.e., $\delta \phi _{\min }=1/\sqrt{N}$) can be understood by the fact that as $\phi \rightarrow 0$, both the variance and the slope of signal for each detection approach zero. In real experiment, however, Cohen \textit{et al.}~\cite{Eisenberg} observed that the peak height decreases exponentially as a function of $N$ and the sensitivity diverges at $\phi =0$~\cite{Bs}. They attributed these observations to the dark counts and the imperfect visibility due to the background counts~\cite{Eisenberg}. In the next section, we investigate the roles of photon loss and phase diffusion noise on the binary-outcome detections.

\section{Binary-outcome detections under the phase diffusion}

We first consider the role of photon loss by introducing a fictitious beam splitter in one of two arms after the phase accumulation~\cite{Enk,Rubin,Dorner,Ma,Escher2011,Zhang}. Only the absorption of photons that carries the phase information is important, so the phase-dependent state can be rewritten as $|\psi (\phi )\rangle =\hat{B}_{T}|\alpha e^{-i\phi }/\sqrt{2}\rangle _{a}\otimes |\alpha /\sqrt{2}\rangle _{b}\otimes |0\rangle _{E}$, where the beam splitter $\hat{B}_{T}$ couples the interferometric mode $a$ and the environment mode $E$ with photon transmission rate $T$~\cite{Bs,Gerrybook}. Tracing over the environment mode, one obtains the phase-dependent state
\begin{equation*}
\hat{\rho}(\phi )=\left\vert \frac{\alpha \sqrt{T}e^{-i\phi }}{\sqrt{2}},\frac{\alpha \sqrt{T}}{\sqrt{2}}\right\rangle \left\langle \frac{\alpha \sqrt{T}e^{-i\phi }}{\sqrt{2}},\frac{\alpha \sqrt{T}}{\sqrt{2}}\right\vert ,
\end{equation*}%
where the transmission rate $T=1$ means no loss and $T=0$ corresponds to a complete photon loss. Comparing with the noiseless case, one can find that
the photon loss leads to a replacement $\alpha \rightarrow \alpha \sqrt{T}$ (i.e., $N\rightarrow NT$) in the output signals. Such a trivial influence cannot explain the imperfect visibility observed by Cohen \textit{et al.}~\cite{Eisenberg}.

Next, we consider the phase-diffusion process after the phase accumulation, which produces a phase fluctuation in one of the two paths (as depicted by Fig.~\ref{fig1}). Formally, the presence of phase noise can be modeled by the following master equation~\cite{Qasimi,Teklu,Liu,Brivio,Genoni11,Genoni12,Escher12,Zhong,Bardhan}: $\partial \hat{\rho}/\partial t=\Gamma _{p}(2\hat{N}_{a}\hat{\rho}\hat{N}_{a}-\hat{N}_{a}^{2}\hat{\rho}-\hat{\rho}\hat{N}_{a}^{2})$, with $\hat{N}_{a}=\hat{a}^{\dag }\hat{a}$ and the phase-diffusion rate $\Gamma_{p}$. Following Refs.~\cite{Genoni12,Brivio}, the solution of $\hat{\rho}$ is given by an integration $\hat{\rho}_{\gamma }(\phi )\propto \int_{\mathbb{R}}d\xi e^{-\xi^{2}/(4\gamma)}\hat{U}(\xi )\hat{\rho}(\phi )\hat{U}^{\dag}(\xi )$, where $\gamma=\Gamma_{p}t$ is a dimensionless diffusion rate. Note that for the noiseless case, the phase-encoded state $\hat{\rho}(\phi)$ obeys $\hat{U}(\xi)\hat{\rho}(\phi)\hat{U}^{\dag}(\xi)=\hat{\rho}(\xi+\phi)$. Replacing $\xi\rightarrow\xi-\phi$ and performing the second beam-splitter operation to $\hat{\rho}_{\gamma}(\phi)$, we obtain the final state,
\begin{equation*}
\hat{\rho}_{\gamma ,\mathrm{out}}(\phi )\!\!=\!\!\int_{\mathbb{R}}\!d\xi \frac{1}{\sqrt{4\pi \gamma }}\exp \left[ -\frac{(\xi -\phi )^{2}}{4\gamma }\right] \hat{\rho}_{\mathrm{out}}(\xi ),
\end{equation*}
where $\hat{\rho}_{\mathrm{out}}(\phi)$ has been given by Eq.~(\ref{output0}). For a very weak diffusion rate (i.e., $\gamma\rightarrow 0$), using $\mathrm{lim}_{\gamma\rightarrow 0}e^{-(\xi-\phi)^{2}/(4\gamma)}/\sqrt{4\pi\gamma}=\delta (\xi-\phi)$, one can easily obtain the final state $\hat{\rho}_{\gamma, \mathrm{out}}(\phi)=\hat{\rho}_{\mathrm{out}}(\phi)$, recovering the noiseless case.

Under the phase diffusion, all the relevant quantities, such as the output signal and the conditional probabilities, can be obtained by integrating the Gaussian with the quantities without diffusion. For instance, the binary-outcome homodyne detection gives the output signal
\begin{equation}
{\langle \hat{p}_{+}(\phi )\rangle }_{\gamma }\!\!=\!\!\int_{\mathbb{R}}\!d\xi \frac{1}{\sqrt{4\pi \gamma }}\exp \left[ -\frac{(\xi -\phi )^{2}}{4\gamma }\right] \langle {\hat{p}_{+}}(\xi )\rangle,  \label{rhoout-2HD}
\end{equation}
where $\langle{\hat{p}_{+}}(\xi)\rangle$ has been given by Eq.~(\ref{p-homody}). Integrating it with the Gaussian, one can obtain the exact numerical result of the output signal [see Fig.~\ref{fig2}(a)], as well as the conditional probability $P(p|\phi)$ for detecting the phase quadrature $p=0$. Due to $\hat{p}_{+}^{2}=\hat{p}_{+}$, one can also obtain the variance $(\Delta {\hat{p}_{+}})^{2}={\langle\hat{p}_{+}\rangle}_{\gamma}(1-{\langle \hat{p}_{+}\rangle}_{\gamma})$ and hence the phase sensitivity $\delta \phi_{H}$ [see Fig.~\ref{fig2}(b)]. According to Eq.~(\ref{sensitivity}), the phase sensitivity can also saturate the CR bound over the whole phase interval. The numerical results in Fig.~\ref{fig2} show that the phase diffusion degrades the fringe resolution and the best sensitivity (see the green dot-dashed lines). In real experiment~\cite{Andersen}, however, the exact results of the signal and the sensitivity without any noise agree very well with their experimental data, implying that the diffusion rate is very small (e.g., $\gamma\lesssim 10^{-4}$). One can also see this from the blue dashed lines of Fig.~\ref{fig2}; the numerical results for $\gamma=10^{-4}$ almost merge with that of the noiseless case. Even for such a small diffusion rate, the phase diffusion has a dramatic influence in the binary-outcome photon counting measurements (see below).

\begin{figure}[t]
\begin{center}
\includegraphics[width=0.9\columnwidth]{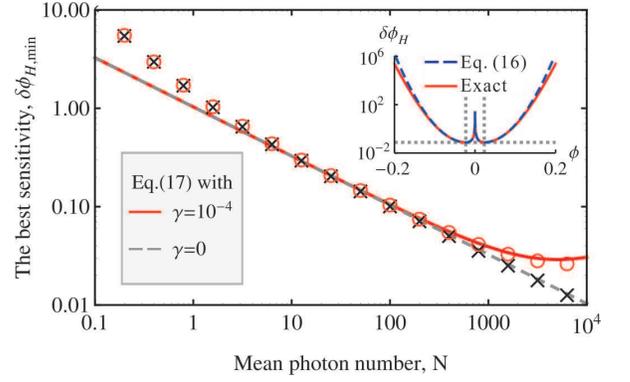}
\caption{(Color online) Log-log plot of the best sensitivity for the binary-outcome homodyne detection with $\gamma =0$ (crosses) and $\gamma =10^{-4}$ (open circles). The open circles and the crosses are obtained by numerically integrating Eq.~(\ref{rhoout-2HD}). Solid (dashed) line: analytical result of $\delta\phi_{H,\min }$, given by Eq.~(\ref{bestsensitivity-2HD}). Inset: the exact and the analytical results of $\delta\phi_H$ as a function of $ \phi $ (in units of $1/\pi$) for $N=200$ and $\gamma =10^{-4} $. Gray dotted lines at $\phi=\pm\Delta/\sqrt{N}$ indicate the location of the best sensitivity. }
\label{fig4}
\end{center}
\end{figure}

To present a unified description of the above measurements, we first analyze the role of phase noise in the binary-outcome homodyne detection. Without any noise, the signal can be approximated as $\langle{\hat{p}_{+}}({\phi})\rangle\approx\sqrt{2/\pi}\exp(-N{\phi}^{2}/2)$. Inserting it into Eq.~(\ref{rhoout-2HD}), we obtain the output signal
\begin{equation}
{\langle \hat{p}_{+}(\phi)\rangle}_{\gamma}\approx\sqrt{\frac{2}{\pi\Delta^{2}}}\exp\left(-\frac{N\phi ^{2}}{2\Delta^{2}}\right) , \label{appro-2HD}
\end{equation}
where we have introduced $\Delta\equiv\sqrt{1+2N\gamma}>1$. Clearly, the phase diffusion leads to the degradation of the fringe resolution, due to the $\mathrm{FWHM}\approx 2\Delta \sqrt{(2\ln 2)/N}$. In addition, we obtain the sensitivity
\begin{equation}
\delta \phi _{H}\approx \frac{\Delta ^{2}}{N|\phi |}\sqrt{\Delta \sqrt{\frac{\pi }{2}}e^{N\phi ^{2}/(2\Delta ^{2})}-\,1},  \label{sensitivity-2HD}
\end{equation}
which shows a good agreement with the exact numerical result at $\phi\sim0$. By maximizing the slope of signal $|\partial{\langle\hat{p}_{+}\rangle}_{\gamma}/\partial\phi|$, we further obtain the optimal phase shift $N\phi_{\mathrm{\min}}^{2}\approx\Delta^{2}$ or $\phi_{\mathrm{\min}}\approx\pm\Delta/\sqrt{N}$ (see the inset of Fig.~\ref{fig4}), which is valid for a small diffusion rate $\gamma$ and large enough $N$. Therefore, from Eq.~(\ref{sensitivity-2HD}) we have
\begin{eqnarray}
\delta \phi _{H,\min }\!\! &\approx &\frac{\Delta (\Delta \sqrt{e\pi /2}-{1})^{1/2}}{\sqrt{N}}  \notag \\ &\!\!=\!\!&\frac{\eta }{\sqrt{N}}\left\{ 1+\frac{3\eta ^{2}+1}{2\eta ^{2}} N\gamma +O[(N\gamma )^{2}]\right\} ,  \label{bestsensitivity-2HD}
\end{eqnarray}
where, for brevity, we introduce $\eta\equiv(\sqrt{e\pi/2}-{1)}^{1/2}\approx 1.03$. Note that for the noiseless case, i.e., $\gamma=0$, the best sensitivity is simply given by $\eta/\sqrt{N}$, recovering Eq.~(\ref{delmin_2HD}). As shown by Fig.~\ref{fig4}, one can find that for large enough $N$ ($\gtrsim 10$), the series expansion of $\delta\phi_{H,\min}$ up to order of $(N\gamma)^1$ works well to predict the best sensitivity (the crosses and the open circles).

Next, we perform similar analysis to the binary-outcome photon counting detections. For the parity detection, the signal under the phase diffusion is given by
\begin{equation}
\langle \hat{\Pi}\rangle _{\gamma }=\int_{\mathbb{R}}\!d\xi \frac{e^{-(\xi -\phi )^{2}/4\gamma }}{\sqrt{4\pi \gamma }}\langle \hat{\Pi}(\xi )\rangle
\approx \frac{1}{\Delta }e^{-N\phi ^{2}/(2\Delta ^{2})},  \label{Pgamma}
\end{equation}
where we have approximated $\langle \hat{\Pi}(\xi)\rangle\approx \exp(-N\xi^{2}/2)$ and introduced $\Delta=\sqrt{1+2N\gamma}$ as done before. For the $Z$ detection, the signal reads
\begin{equation}
\langle \hat{Z}\rangle _{\gamma }=\int_{\mathbb{R}}\!d\xi \frac{e^{-(\xi -\phi )^{2}/4\gamma }}{\sqrt{4\pi \gamma }}\langle \hat{Z}(\xi )\rangle \approx \frac{1}{\Delta _{0}}e^{-N\phi ^{2}/(2\Delta _{0})^{2}}, \label{Zgamma}
\end{equation}
where $\Delta_{0}\equiv \sqrt{1+N\gamma}$. Similar to the binary-outcome homodyne detection, the fringe resolution of each detection degrades by a factor $\Delta $ or $\Delta _{0}$, as the $ \mathrm{FWHM}\approx 2\Delta \sqrt{(2\ln 2)/N}$ for the parity detection and $4\Delta_{0}\sqrt{(\ln 2)/N}$ for the $Z$ detection. Actually, the role of phase diffusion in the above three measurements is the same; it is uniquely determined by the product $N\gamma$, instead of $\gamma$ or $N$ alone. When $N\gamma\ll 1$, the diffusion is negligible and the resolution $\Delta x\sim \lambda /(2\sqrt{N})$. With the increase of $N\gamma$, the resolution degrades and its scaling undergoes a transition from $ N^{-1/2}$ to $N^{0}$, due to the $\mathrm{FWHM}\rightarrow 4\sqrt{\gamma\ln 2}$ as $N\gamma\gg 1$.

\begin{figure}[hptb]
\begin{center}
\includegraphics[width=0.9\columnwidth]{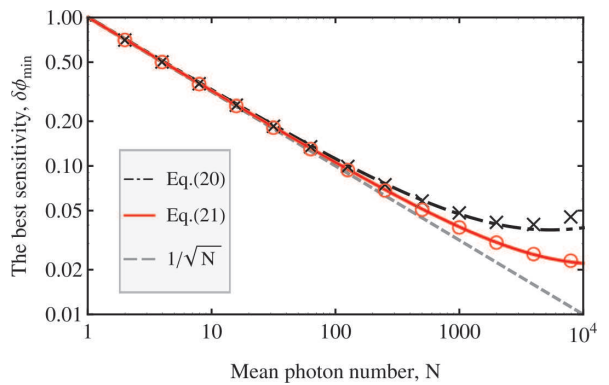}
\caption{(Color online) Log-log plot of the best sensitivities for the parity measurement (crosses) and the $Z$ detection (open circles) with a given phase-diffusion rate $\gamma =10^{-4}$. Dot-dashed (red solid) line: the analytical result of $\delta \phi _{\Pi ,\min }$ ($ \delta \phi _{Z,\min }$), given by Eqs.~(\ref{sen_P}) and (\ref{sen_Z}), respectively. Dotted line: the shot-noise limit $\delta \phi _{\min }=1/\sqrt{N}$. The exact results, indicated by the crosses and the open circles, are obtained by numerically integrating the output signals for the two measurement schemes.}
\label{fig5}
\end{center}
\end{figure}

Unlike the homodyne detection, the phase diffusion has a dramatic influence on the sensitivity for the binary-outcome photon counting. Even for a small dephasing rate $\gamma \sim 10^{-4}$, one can find that the peak heights $ 1/\Delta $ and $1/\Delta _{0}$ decrease as $N$ increases, similar to Ref.~\cite{Eisenberg}. The degradation of the peak heights results in nonzero variance at $\phi=0$, while the slope of signal tends to zero as $\phi \rightarrow 0$. Therefore, the sensitivities of the two detections diverge at the phase origin. As shown in Fig.~\ref{fig3}, one can find that the best sensitivity occurs at $\phi \neq 0$. A similar result has been observed by Cohen \textit{et al.}~\cite{Eisenberg}. To understand this behavior, we now calculate the best sensitivity by minimizing analytical result of $\delta\phi$. For the parity measurement, we find that the best sensitivity occurs at $\phi _{\min }\approx \pm \Delta \sqrt{\lbrack 1+w(-e^{-1}\Delta ^{-2})]/N}$, where $w(z)$ denotes the Lambert W function (also called the product logarithm), defined by the principal solution of $w$ in the equation $z=we^{w}$. With the optimal phase $\phi_{\min}$, we obtain the best sensitivity,
\begin{eqnarray}
\delta \phi _{\Pi ,\min } &\approx &\frac{1}{\sqrt{N}}\Delta ^{2}\exp \left[ \frac{1+w\left( -e^{-1}\Delta ^{-2}\right) }{2}\right]  \notag \\ &\approx &\frac{1}{\sqrt{N}}\left( 1+\sqrt{N\gamma }+\frac{11N\gamma }{6} \right) ,  \label{sen_P}
\end{eqnarray}
where, in the last step, we have expanded $\delta \phi _{\Pi ,\min }$ up to order of $(N\gamma )^{1}$. Note that the first line of Eq.~(\ref{sen_P}) is valid in the parameter regime $z=e^{-1}\Delta ^{-2}\in (0,e^{-1})$, for which the W function obeys the inequality $0<-w(-z)<1$ and hence $1+w(-z)>0$. The second line of Eq.~(\ref{sen_P}), as shown by the dot-dashed curve of Fig.~ \ref{fig5}, agrees quite well with the exact numerical result (the crosses).

Similarly, for the $Z$ detection, we obtain the best sensitivity $\delta\phi_{Z,\mathrm{\min}}\approx \Delta_{0}/\sqrt{-Nw(-e^{-1}\Delta_{0}^{-1})}$, which can be further approximated as
\begin{equation}
\delta \phi _{Z,\mathrm{\min }}\approx \frac{1}{\sqrt{N}}\left( 1+\frac{ \sqrt{N\gamma }}{2}+\frac{17N\gamma }{24}\right) .  \label{sen_Z}
\end{equation}
From Fig.~\ref{fig5}, one can find that for $N\lesssim 10^{2}$ and $\gamma\sim 10^{-4}$ (i.e., $N\gamma\lesssim 10^{-2}$), the best sensitivities almost follow the shot-noise scaling. When $N\gamma >10^{-2}$, both $\delta\phi_{\Pi, \mathrm{\min}}$ and $\delta\phi_{Z,\mathrm{\min}}$ become worse. However, the $Z$ detection gives a slightly better sensitivity than that of the parity detection, which can be easily understood by comparing Eqs.~(\ref{sen_P}) and (\ref{sen_Z}).

Finally, we should point out that our result, Eq.~(\ref{sensitivity}), remains valid even in the presence of the noise. This can be easily found from explicit forms of the variances $(\Delta \hat{\Pi})^{2}=4P(+1|\phi )P(-1|\phi )$ and $(\Delta \hat{Z})^{2}=P(0|\phi )P(\emptyset |\phi )$, where $P(\pm 1|\phi )=(1\pm \langle \hat{\Pi}\rangle _{\gamma })/2$ and $ P(0|\phi )=\langle \hat{Z}\rangle _{\gamma }$ are the conditional probabilities for the parity and the $Z$ detections. Indeed, the phase sensitivity of any binary-outcome detection always saturates the CR bound.

\section{Conclusion}

In summary, we have shown that phase sensitivity with a general binary-outcome measurement always saturates the Cram\'{e}r-Rao bound. Its validity is demonstrated by the recent experiments based on coherent-light Mach-Zehnder interferometer~\cite{Andersen,Eisenberg}. The observed super-resolution in fringe pattern and the shot-noise limited sensitivity can be well understood as suitable data processing over the conditional probabilities $P(p|\phi)$ and $P(n,m|\phi)$ for detecting a phase quadrature $p$ in a homodyne detection and a pair of photons $\{n,m\}$ in a photon counting measurement, respectively.

We consider the performance of the binary-outcome homodyne measurement, the parity, and the $Z$ detections in the presence of phase diffusion. Interestingly, we find that the role of phase diffusion is uniquely determined by a product of the mean photon number $N$ and the diffusion rate $\gamma $. When $N\gamma \ll 1$, both the resolution and the sensitivity almost exhibit the shot-noise scaling $\sim 1/\sqrt{N}$. Except for the experimental imperfections, we show that a very weak phase diffusion can dramatically change the behavior of the sensitivity in the binary-outcome photon counting. Our analytical results also confirm that the $Z$ detection gives a better sensitivity than that of the parity detection, in agreement with the experiment observation~\cite{Eisenberg}.

\begin{acknowledgments}
We thank Professor J. P. Dowling for helpful discussions. This work was supported by the NSFC (Contracts No.~11174028, No.~11274036, and No.~11322542), the MOST (Contract No.~2014CB848701), and the FRFCU (Contract No.~2011JBZ013).
\end{acknowledgments}


\begin{thebibliography}{99}
\bibitem{Bs} Throughout this work, we consider the action of the beam splitter as $\hat{B}_{T}=\exp [-i\theta \hat{J}_{y}^{(k,l)}]$, where $\hat{J}_{y}^{(k,l)}=(\hat{a}_{k}^{\dag}\hat{a}_{l}-$H.c.$)/2i$. The BS couples two optical modes $k$ and $l$ with the photon transmission rate $T$ and the mixing angle $\theta =2\arccos (\sqrt{T})$. For a 50:50 beam splitter, the photon transmission rate $T=1/2$ and $\theta =\pi /2$, corresponding to a $\pi /2$ pulse in atomic spectroscopy. If one adopts another kind of the beam splitter, $\hat{B}_{1/2}=\exp [-i\pi \hat{J}_{x}^{(k,l)}/2]$, where $\hat{J}_{x}^{(k,l)}=(\hat{a}_{k}^{\dag }\hat{a}_{l}+$H.c.$)/2$, the phase should be shifted as $\phi \rightarrow \phi +\pi $, as shown in Refs.~\cite{Eisenberg,Gerrybook}.
\end{thebibliography}
\end{document}